# An Analytical Evaluation of Matricizing Least-Square-Errors Curve Fitting to Support High Performance Computation on Large Datasets

Poorna Banerjee Dasgupta

*M.Tech Computer Science and Engineering, Nirma Institute of Technology*
*Ahmedabad, Gujarat - India*

***Abstract*** *— The procedure of Least Square-Errors curve fitting is extensively used in many computer applications for fitting a polynomial curve of a given degree to approximate a set of data. Although various methodologies exist to carry out curve fitting on data, most of them have shortcomings with respect to efficiency especially where huge datasets are involved. This paper proposes and analyzes a matricized approach to the Least Square-Errors curve fitting with the primary objective of parallelizing the whole algorithm so that high performance efficiency can be achieved when algorithmic execution takes place on colossal datasets.*

## I. INTRODUCTION

The technique of Least Square-Errors (LSE) curve fitting on data is a standard tool in statistical regression analysis. Figure 1 shows an example of LSE data-fitting with a quadratic function [7].

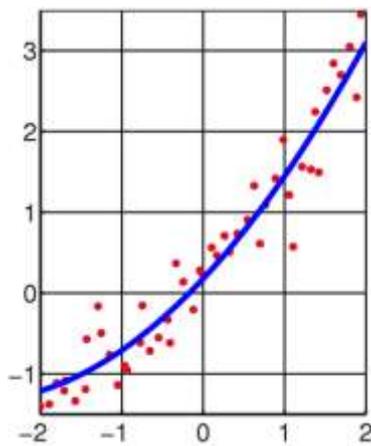

Fig. 1. Fitting data points with a quadratic LSE function

LSE curve fitting on data has been largely deployed in many scientific computer applications – be it determining *Light Transfer Characteristics* of an optical imaging system in a satellite or for *Weather Forecasting*. When such applications involve large datasets, commercially available software algorithms typically slow down the curve fitting process, mostly because inherent parallelism in the input datasets is not fully exploited. This paper suggests and explores a matricized algorithmic approach for parallelizing the LSE curve fitting procedure, in order to achieve high performance efficiency, especially so that the suggested algorithm can be deployed on many-core parallel processors like General Purpose Graphic Processing Units (GPGPUs)[4]. The algorithmic approach described in this paper has been specially formulated for lower-order polynomial curve fitting.

## II. MATRICIZING LEAST SQUARE-ERRORS CURVE FITTING

In order to parallelize and hence improve efficiency of the LSE curve fitting, it is customary that the input data as well as the LSE coefficients be represented in the form of matrices and vectors. This is explained further below:

If the input data-set is represented by pairs of the type $(x_i, y_i)$ where $1 \leq i \leq n$, $n \geq 2$, and $n$ is the number of data-points, then the best fitting curve $f(x)$ has the least square error, i.e.[1][5]:

$\Pi = \sum[y_i - f(x_i)]^2 = minimum$, where $i = 1$ to $n$

For, example if we want to determine a best-fit, LSE straight line on the given set of data, then $f(x)$ will be given by:

$f(x) = a + bx$, where $a$ and $b$ are coefficients to be determined.

Similarly, if we want to determine a second degree best-fit LSE curve, then $f(x)$ will be given by:

$f(x) = a + bx + cx^2$, where $a$, $b$ and $c$ are coefficients to be determined. Similarly, for a $m^{th}$ degree polynomial fit:

$f(x) = a_0 + a_1 x + a_2 x^2 + ... + a_m x^m$, where $a_0, a_1, a_2,..., a_m$ are coefficients to be determined.

To obtain the least square error, the unknown coefficients must yield zero first derivatives, which lead to the following equations:

$\partial \Pi / \partial a_j = 2\sum [y_i - (a_0 + a_1 x_i + a_2 x_i^2 + ... + a_m x_i^m)] = 0$, where $i = 1$ to $n$, $j = 0$ to $m$.

Expanding the above set of equations, we get:

$\sum x_i^j y_i = a_0 \sum x_i^j + a_1 \sum x_i^{j+1} + ... + a_m \sum x_i^{j+m}$, where $i = 1$ to $n$, $j = 0$ to $m$.

The unknown coefficients $a_0, a_1,..., a_m$ can hence be obtained by solving the above set of linear equations.

The set of equations described above indicate that in order to determine the unknown coefficients $a_0, a_1,..., a_m$, we have to solve a system of linear equations of the form **AX = B**, where the matrices **A, X** and **B** are given as shown:

**Matrix X** = [ $a_0$  $a_1$  $a_2$ ... $a_m$ ]
**Matrix B** = [ $\sum y_i$   $\sum x_i y_i$   $\sum x_i^2 y_i$  ...  $\sum x_i^m y_i$ ]
**Matrix A** =

| 1 | $\sum x_i$ | $\sum x_i^2$ ... | $\sum x_i^m$ |
| $\sum x_i$ | $\sum x_i^2$ | $\sum x_i^3$ ... | $\sum x_i^{m+1}$ |
| ... | | | |
| $\sum x_i^m$ | $\sum x_i^{m+2}$ | $\sum x_i^{m+3}$ ... | $\sum x_i^{2m}$ |

The matrix **X** can now be solved by evaluating the inverse of matrix **A**, i.e. **X = A$^{-1}$B**. In this paper's algorithmic implementation for testing accuracy of results, the matrix **X** has been solved for using the method of *Gaussian Elimination*.

### III. ACCURACY ANALYSIS OF RESULTS

After the matricization for LSE curve fitting has been done, its now time to test the accuracy of results produced by the proposed approach. As a standard for comparison, MATLAB's *polyfit()* function has been also used on the same input data for fitting linear, quadratic and cubic curves and the determined coefficients have then been compared. MATLAB's *polyfit()* function uses an indirect method of determining the least-squares coefficients. This method is based on constructing the *Vandermonde* matrix **V** and then performing *QR factorisation* of **V**, where **Q** is an orthogonal matrix and **R** is an upper triangular matrix as shown. The QR factorisation is usually carried out using *Holder Reflections* [2]. Therefore we have:
**V p = Y**
If **V = QR**, where **Q** is an orthogonal matrix, then:
**Q.Q$^T$ = I**, where **T** is transpose operator and **I** is the identity matrix. Substituting these values, we have:
**(QR) p = Y** or **(Q Q$^T$ )R p = Q$^T$Y**
i.e. **p = R$^{-1}$(Q$^T$Y)** or **p = (Q$^T$Y) \ R**, where \ is the special matrix division operator used by MATLAB. Hence, the *polyfit()* function returns the unknown coefficients in the array **p**.
**Matrix p** = [ $a_0$  $a_1$  $a_2$ ... $a_m$ ]
**Matrix Y** = [ $y_1$  $y_2$  $y_3$ ...  $y_n$ ]

For testing purposes, several comparisons were made for different polynomial orders and the coefficients were calculated for each order along with the corresponding least-squares errors for the fitted data. The calculated coefficients, along with *polyfit()'s* coefficients are shown in the following tables. Table 1 shows a sample dataset, Tables 2, 3 and 4 show the calculated coefficients for this dataset for polynomial orders 1, 2 and 3 respectively and Table 5 shows the fitted data with calculated coefficients and the corresponding least squared errors.

**TABLE I**
**SAMPLE DATASET**

| x | y |
|---|---|
| 39.206 | 751.912 |
| 29.74 | 567.121 |
| 21.31 | 403.746 |
| 12.087 | 221.738 |
| 1.812 | 18.8418 |
| 0.001 | 1.88672 |

**TABLE II**
**ORDER 1 COEFFICIENTS FOR BEST-FIT LSE CURVE**

| Generated Values | *polyfit()* Values |
|---|---|
| $a_0$=-8.356 | $a_0$=-8.356 |
| $a_1$=19.3496 | $a_1$=19.3496 |
| R = 0.9997 | |

**TABLE III**
**ORDER 2 COEFFICIENTS FOR BEST-FIT LSE CURVE**

| Generated Values | *polyfit()* Values |
|---|---|
| $a_0$= -6.5106 | $a_0$= -6.5109 |
| $a_1$= 18.8735 | $a_1$= 18.8735 |
| $a_2$ = 0.0127 | $a_2$ = 0.0127 |
| R = 0.9998 | |

**TABLE IV**
**ORDER 3 COEFFICIENTS FOR BEST-FIT LSE CURVE**

| Generated Values | *polyfit()* Values |
|---|---|
| $a_0$= -4.7553 | $a_0$= -4.7551 |
| $a_1$= 17.5105 | $a_1$= 17.5109 |
| $a_2$ = 0.1086 | $a_2$ = 0.1086 |
| $a_3$ = -0.0016 | $a_3$ = -0.0016 |
| R = 0.9996 | |

**TABLE V**
**FITTED DATA WITH ORDER 3 COEFFICIENTS**

| y | $y_f$ =f(x) | $y_p$=$f_p$(x) | $e_f$=y-$y_f$ | $e_p$=y-$y_p$ |
|---|---|---|---|---|
| 751.912 | 751.18396 | 752.285156 | 0.728027 | -0.37317 |
| 567.121 | 569.500305 | 569.985718 | -2.37933 | -2.86475 |
| 403.746 | 402.053284 | 402.235626 | 1.69272 | 1.510376 |
| 221.738 | 219.903793 | 219.939758 | 1.83421 | 1.798248 |
| 18.8418 | 27.321678 | 27.321703 | -8.47988 | -8.4799 |
| 1.88672 | -4.736779 | -4.737589 | 6.6235 | 6.624309 |
| | | | $\sum e_f^2$ = 128.1999 | $\sum e_p^2$ = 129.6512 |

**R** represents the *Correlation Coefficient* in Tables 2, 3 and 4. In Table 5, the function *f(x)* represents the best-fit polynomial with generated coefficients. The function *$f_p(x)$* represents the best-fit polynomial with MATLAB's *polyfit()* coefficients. From Table 5, the sum of Least Square errors were calculated, i.e.:
$\sum e_f^2$ = 128.199937, $\sum e_p^2$ = 129.651164.

According to the definition, the best-fit curve is the one which yields minimum least squared errors. It can be seen that the generated coefficients yield a lower least-squares error than the *polyfit()* coefficients and hence a best-fit curve is produced as compared to *polyfit()*. This procedure was repeated for several other data-sets, and consistent results were obtained.

## IV. PERFORMANCE SPEED-UP ANALYSIS

As mentioned in the previous sections of this paper, the primary objective of Matricizing the LSE curve fitting procedure is to achieve higher efficiency and speed-up during execution, by exploiting parallelism. Many-core processors like General Purpose Graphic Processing Units (GPGPUs) and programming languages like CUDA are specially designed to parallely operate on datasets represented in the form of vectors and matrices [3][6]. Datasets involved in LSE curve fitting are hitherto in the form of vectors involving two or more variables and after matricizing the LSE equations, the process of curve fitting becomes ideal for implementation on parallel processors. For testing purposes, one such implementation has been done in the scope of this paper where the programming language has been chosen to be CUDA and the parallel platform has been chosen as NVIDIA *Quadro 4000* with compute capability 2.0 and 256 cores. It has been found that even with a dataset having thousands of data-points, speed-ups of the order of ~100 can be achieved compared to the sequential execution of the same on any normal multi-core processor.

## V. CONCLUSIONS & FUTURE SCOPE OF WORK

The procedure of curve fitting based on Least Square Errors has many scientific computer applications and involves fitting a polynomial function that best approximates a given set of data-points. Such a best-fit polynomial is then described in terms of polynomial coefficients. This paper proposes and elucidates how to matricize the entire procedure to exploit parallelism and obtain higher execution speed, especially where colossal datasets are involved. In order to qualitatively assess the precision of the results obtained, a comparative analysis with MATLAB's *polyfit()* function has been done, and minimum least square errors and hence a best-fit curve has been obtained.

As further extension to the research work carried out in this paper, methods other than *Gaussian elimination* can be carried out for calculating the inverse matrices and determinants. Also the analysis can be further extended for higher order polynomials as well.

## AUTHOR'S PROFILE

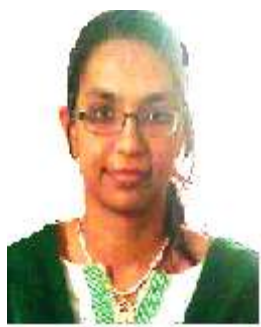

**Poorna Banerjee Dasgupta** has received her B.Tech & M.Tech Degrees in Computer Science and Engineering from Nirma Institute of Technology, Ahmedabad, India. She did her M.Tech dissertation at Space Applications Center, ISRO, Ahmedabad, India and has also worked as Assistant Professor in Computer Engineering dept. at Gandhinagar Institute of Technology, Gandhinagar, India from 2013-2014 and has published research papers in reputed international journals. Her research interests include image processing, high performance computing, parallel processing and wireless sensor networks